# Separating hyperfine from spin-orbit interactions in organic semiconductors by multi-octave magnetic resonance using coplanar waveguide microresonators

G. Joshi[1], R. Miller[1], L. Ogden[1], M. Kavand[1], S. Jamali[1], K. Ambal[1], S. Venkatesh[2], D. Schurig[2], H. Malissa[1], J.M. Lupton[1,3], and C. Boehme[1,a]

1) *Department of Physics and Astronomy, University of Utah, Salt Lake City, UT 84112, USA*

2) *Department of Electrical and Computer Engineering, University of Utah, Salt Lake City, UT 84112, USA*

3) *Institut für Experimentelle und Angewandte Physik, Universität Regensburg, D-93040 Regensburg, Germany*

(Dated: 02 February 2016)

Separating the influence of hyperfine from spin-orbit interactions in spin-dependent carrier recombination and dissociation processes necessitates magnetic resonance spectroscopy over a wide range of frequencies. We have designed compact and versatile coplanar waveguide resonators for continuous-wave electrically detected magnetic resonance, and tested these on organic light-emitting diodes. By exploiting both the fundamental and higher-harmonic modes of the resonators we cover almost five octaves in resonance frequency within a single setup. The measurements with a common π-conjugated polymer as the active material reveal small but non-negligible effects of spin-orbit interactions, which give rise to a broadening of the magnetic resonance spectrum with increasing frequency.

---

a) boehme@physics.utah.edu

Multi-frequency electron paramagnetic resonance (EPR) [1] is an important tool for resolving the different interactions of an electron spin whose influences on magnetic resonance spectra have distinct dependencies on the applied magnetic field strength. For example, zero-field splitting and nuclear hyperfine interactions are independent of field and therefore dominate the resonance characteristics in the low-field regime. On the other hand, $g$-factors and anisotropies thereof arise due to spin-orbit interactions (SOI) and are best resolved at higher fields since the associated Zeeman energy shifts scale directly with magnetic field strength. Conventional multi-frequency EPR experiments are usually only carried out at a few selected frequencies for which resonators and microwave bridges are readily available. However, since spin polarization is measured, these experiments are limited by the Boltzmann factor determining thermal spin polarization, which decreases with decreasing Zeeman splitting and increasing temperature. In contrast, in electrically detected magnetic resonance (EDMR) [2] these limitations play a lesser role because the signal is obtained from sample conductivity rather than from the transmitted or reflected microwave intensity, simplifying the microwave circuitry required considerably. Conductivity depends solely on singlet-triplet spin-permutation symmetry of a carrier pair rather than on spin polarization [3, 4, 5, 6, 7], so that EDMR can be observed at static fields as weak as a few mT [5, 6, 8]. In addition to the actual microwave resonator, EDMR only requires a suitable microwave source and waveguide, but does not require a microwave detector.

Planar waveguide resonators are used extensively in different areas of physics research such as in superconducting quantum bits [9], circuit quantum electrodynamics [10], broadband radiation detectors [11], and magnetic resonance spectroscopy [12, 13, 14]. However, most of the work described previously relates to superconductors at low temperatures and employs only the fundamental frequency of the resonator. Here, we exploit the fundamental and higher harmonic modes of coplanar waveguide (CPW) resonators for EDMR spectroscopy at room temperature. In a conventional EPR spectrometer the use of higher harmonics of a resonator is technically challenging because of the nonlinear properties of microwave detectors over a wide frequency range, requiring a separate calibration for each frequency – a limitation that is circumvented in EDMR.

We constructed a series of quarter-wave resonators spanning the frequency range of 1-6 GHz in nominally equidistant steps. The resonators were patterned on a commercial microwave circuit board (AD1000, Arlon Microwave Materials) of dimensions $1.27 \times 3.81$ cm$^2$. The resonators are capacitively coupled to a feed line on the open end with the opposite end shorted. The length of each resonator and the dielectric constant ($\varepsilon_r \approx 10$) of the substrate determine the fundamental resonator frequency $f_0$. In order to estimate the amplitude $B_1$ of the microwave field above the plane of the resonator where the OLED sample is placed, we simulated the characteristics of a cross-section of the CPW using the COMSOL Multiphysics software package. A particular concern was the weakening and distortion of the field due to the metallic electrodes of the OLED. We took this problem into account by adding a 100 nm Al layer at a distance of 20 µm from the CPW as illustrated by the horizontal red line in Fig. 1(a), which shows the resulting $B_1$ amplitude for excitation at 1 W microwave power at 1 GHz frequency (note that the Al contact is part of the OLED device structure, labelled "OLED" in Fig. 1(a), whose thickness is negligible on the displayed scale). The simulation implies that spin resonance can be excited even though the aluminum electrode attenuates the microwave power. All resonators were fabricated with a width $w = 152$ µm of the center conductor and a slot width $g = 76$ µm. These dimensions were chosen to give a 50 Ω impedance for the given substrate material with $\varepsilon_r = 10$.

Figure 1(b) shows a sketch of the entire setup with a close-up photograph of the CPW given in panel (c). The CPW is coupled to the feed line by a segment of length 2.54 mm lying parallel to the feed line at an edge-to-edge distance of 200 µm. This arrangement forms a constant-coupling capacitor with a resonance $Q$-factor <100. To accommodate the length of the resonator on an area which corresponds to active device areas of OLEDs, the CPW is meandered with a center-to-center distance between segments of ~1 mm, sufficiently large to prevent cross-talk. The feed line ends in an end launch connector (Southwest Microwave Inc.) to enable physical connection to a high-frequency coaxial cable. Finally, ground stitching with an array of vertical interconnects through the circuit board, spaced at sub-wavelength distances along the CPW, ensures that the CPW edges are pinned to a defined potential.

Although the CPW will operate free standing, we carried out the measurements in a standard Oxford Instruments CF935 continuous-flow cryostat mounted in the magnet of a Bruker ElexSys E580 X-band spectrometer to offer atmospheric and temperature control along with high accuracy in the applied external magnetic field $B_0$. The corresponding probe head and sample holder designed to enable these experiments are shown in Figure 1(d). The circuit board containing the resonator is mounted on a rigid brass support plate together with the end launch connectors to connect to the microwave source through an SMA cable. The sample holder assembly is attached to the support plate in such a way that the active region of the OLED lies on top of the wound-up resonator. A Helmholtz coil is placed above and below the resonator to allow acquisition of EDMR spectra through modulation of the Zeeman ($B_0$) field, as shown in Figure 1(e).

The individual resonator probe heads are characterized by measuring the reflection coefficients ($S_{11}$) as a function of frequency. Figure 2 shows the reflection spectra of a resonator exhibiting a fundamental frequency of $f_0 = 2.33$ GHz. The harmonics are given by $f_n = (2n + 1) f_0$ (with $n$ an integer labelling the higher harmonic) and are marked in the figure by vertical lines. The slight deviation of the measured data from this scaling law at higher frequencies is attributed to the dependence of dielectric constant on frequency, which also affects the coupling capacitance.

EDMR was carried out with an OLED based on the π-conjugated polymer poly[2-methoxy-5-(2-ethylhexyloxy)-1,4-phenylenevinylene] (MEH-PPV). The layer structure and the fabrication of these OLEDs in the context of EDMR has been described previously [6]. A constant bias of ~3 V is applied to the OLED resulting in a device current of 20 µA. The resonator is connected to an Agilent broad-frequency analog signal generator (model EXG N5173B). A lock-in detection scheme is employed to acquire the EDMR spectra by modulating the external Zeeman field $B_0$ with a field $B_m$ at 500 Hz. Changes in the OLED current under resonance are detected using an SR570 low-noise current preamplifier (Stanford Research

Systems), the voltage output of which is lock-in detected using the continuous wave signal input channel of the EPR spectrometer. The modulation amplitude $B_m$ of the Helmholtz coil is calibrated in the usual way [15, 16] by recording EDMR spectra in the overmodulated regime where the observed line width scales with $B_m$. A modulation amplitude of $B_m \approx 0.3$ mT results in the strongest EDMR signal while avoiding artificial line broadening.

Figure 3(a) shows a collection of field-modulated EDMR spectra obtained with six different CPW resonators operating at the fundamental and higher-harmonic frequencies. The six resonators are coded in different colors. The graph plots the differential current change $(\delta I / \delta B) B_m$. Generally, for measurements at higher frequencies a substantial reduction in EDMR amplitude was found, which is attributed to increased inhomogeneity in $B_1$ field distribution, a reduction in quality factor and impeded resonator coupling above the fundamental frequency [17]. The EDMR signal itself results from the change of OLED current arising from microwave induced transitions between singlet and triplet spin states of electron-hole pairs. These transitions modify the pairs' recombination and dissociation [18, 19, 3, 20, 21]. Since two resonant species are involved in EDMR, and each species experiences a Gaussian broadening of the resonance due to hyperfine interactions with local nuclear magnetic moments [22], the overall EDMR line shape can be described by the derivative of the sum of two Gaussians. Figure 3(b) compares normalized EDMR spectra recorded at the highest and lowest $B_0$ fields (resonance frequencies), along with double Gaussian fits (black lines). The high-frequency spectrum shows substantial broadening. While the agreement between data and the model is good for low microwave frequencies, it becomes progressively worse as the frequency increases. This indicates that the double Gaussian model is not completely adequate, and other frequency dependent line shape contributions do play a role at higher frequencies. However, in the frequency range considered in this work, the double Gaussian model describes the overall line shape reasonably well. To extract the parameters governing this frequency-dependent broadening we performed a global fit to a function $Y'$ of magnetic field $B_0$ and microwave frequency $f$ of the form

$$Y'(f,B) = A\frac{(f/\gamma)-B}{\sigma_1^3}e^{-\frac{(B-f/\gamma)^2}{2\sigma_1^2}} + \frac{A}{R}\frac{(f/\gamma)-B}{\sigma_2^3}e^{-\frac{(B-f/\gamma)^2}{2\sigma_2^2}}. \quad (1)$$

Here, R is the ratio between the areas A of two Gaussians of width $\sigma_1$ and $\sigma_2$. These widths consist of both a constant term and a contribution linearly dependent on magnetic field

$$\sigma_i = \sqrt{\Delta B_{hyp,i}^2 + \alpha_i^2 B^2} \quad (2)$$

with $i \in \{1,2\}$. In our model, we assume that the field-independent contribution to the line width $\Delta B_{hyp,i}$ originates from hyperfine coupling [22, 5]. In addition, the spin-orbit interaction leads to a contribution to the resonance line width due to local variations in the g-factor. This spread $\Delta g$ induces inhomogeneous line broadening which is proportional to the external magnetic field $B_0$, with $\alpha_i$ being the spin-orbit induced line broadening parameter that is responsible for the $\Delta g$-effect. In the fit algorithm, all measured spectra are described by the same parameters $\Delta B_{hyp,i}$ and $\alpha_i$.

Figure 3(c) plots the extracted widths of the two Gaussians as a function of magnetic field, showing a clear broadening with increasing field. The plot exhibits plateaus at low fields that correspond to the hyperfine field distributions. The asymptotic linear increase of the widths at higher fields is described by the slopes $\alpha_i$. To obtain a reliable estimate of the confidence level of the parameters we utilized a "bootstrap method" [23] which is suitable for such nonlinear models. In brief, artificial data sets were created using the fit result, with the same noise level as the original data. These data sets were in turn fit again with the model, providing a value for the spread of the fit parameter. The resulting parameters and errors are $\alpha_1 = 1.78(13) \times 10^{-4}$, $\alpha_2 = 4.82(1.04) \times 10^{-4}$, $\Delta B_{hyp,1} = 0.1909(3)$ mT and $\Delta B_{hyp,2} = 0.7227(26)$ mT, with the Gaussian area ratio $R = 0.3084(13)$. Note that the fact that this value deviates from unity indicates the presence of an additional spin-dependent process which involves one of the polaron states that are represented fit resonances. The nature of this additional process is different from the simple electron-hole pair model. A further contribution to spin-dependent conductivity could result from the interaction with trapped charges [24], or through the interaction of charges with triplet excitons [25]. Knowledge of the errors on the fit parameters allows the extrapolation of the field dependence of the line widths to high magnetic fields. The blue and red lines in the shaded regions in Fig. 3(c) show the expected spectral widths of the narrow and broad resonance peak, respectively, while the shaded regions themselves represent a 95% confidence intervals. Since the contributions from the field-independent line width (the hyperfine interactions) are significant over the entire frequency range explored, measurements at higher frequencies will be needed to provide improved differentiation of the field-dependent line-width contributions. We note that earlier OLED EDMR measurements [5] inferred a steeper increase of line width as a function of frequency. While the extracted low-frequency plateau was similar (22% lower than the present result), the largest width measured at 9.6 GHz, using pulsed EDMR, was found to be 110% larger than the value reported here. We conclude that in absence of a global fit of multi-frequency data as presented in this study, the double Gaussian fit of an EDMR spectrum at a single frequency leads to large uncertainties on the widths and the peak ratios R of such spectra. For the 9.6GHz EDMR spectrum of our earlier work, this caused an overestimation of the smaller peak width and an associated underestimation of the effects of power broadening [26] for the observed resonance line.

In conclusion, we have developed a series of CPW resonators to study the EDMR line width of OLEDs over a wide range of microwave frequencies. Using this approach, we can distinguish between field-independent and field-dependent contributions to the resonance width of electrons and holes. These contributions result from hyperfine and spin-orbit interactions, respectively. The results unambiguously show that spin-orbit coupling can play a role in organic semiconductors despite the low atomic-order number of the constituent elements, and imply that spin-orbit coupling likely contributes to magnetoresistive effects above fields of ~100 mT [27, 28].

**Acknowledgements**

This work was supported by the US Department of Energy, Office of Basic Energy Sciences, Division of Materials Sciences and Engineering under Award #DE-SC0000909. L.O. and K. A. acknowledge support for their contributions through the Utah MRSEC center, grant #1121252.

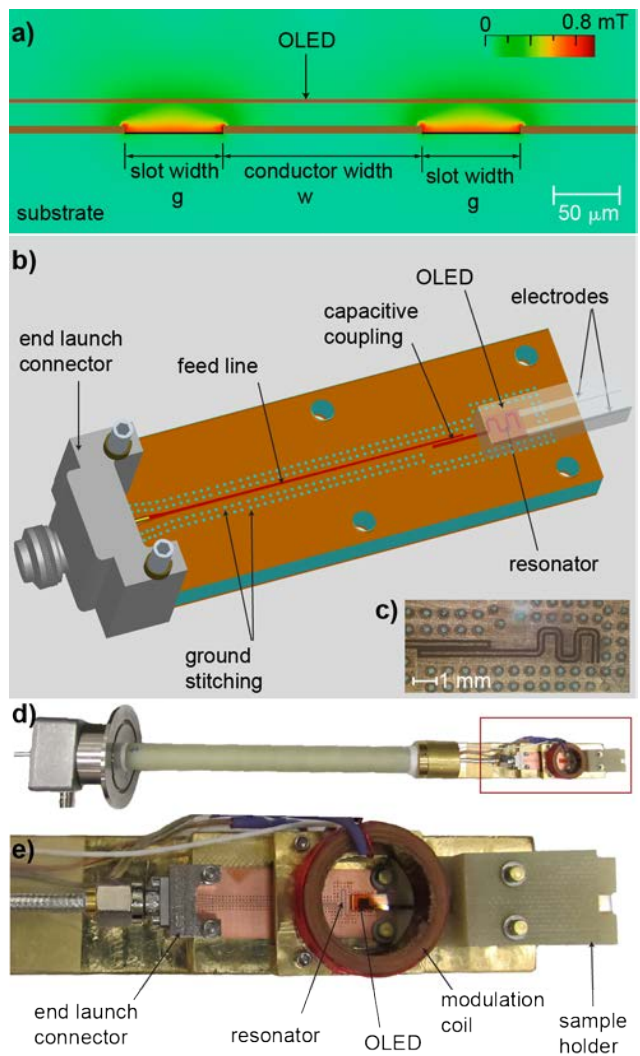

**Figure 1:** (color online) **Design and implementation of CPW resonators for multi-frequency EDMR spectroscopy of OLEDs.** **(a)** Simulation of the distribution of the oscillating magnetic field amplitude $B_1$ with an OLED (active layer and electrode) placed 20 µm above the waveguide. **(b)** Sketch of the CPW resonator illustrating the circuit-board layout with an end-launch connector and an OLED device placed on top. **(c)** Photograph of the circuit board showing the resonator and the coupling capacity between feed line and CPW. **(d)** Image of the entire EDMR probe-head cryostat insert. (e) Image of the resonator circuit board integrated in the probe-head setup showing the $B_0$ modulation coil, the sample holder and the contact system.

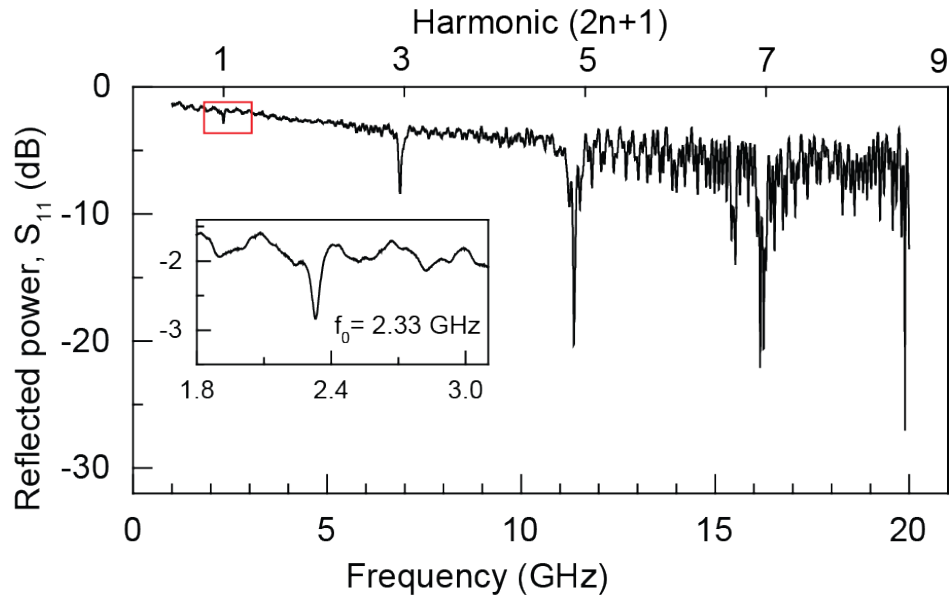

**Figure 2: Reflection coefficient function S11 of a CPW mounted on the probe head.** The fundamental frequency is determined to be $f_0$ =2.33 GHz with harmonics at $f_n = (2n + 1) f_0$ with $n$=1, 2, 3, 4, marked by vertical ticks of the upper scale. Slight deviations of the observed resonances from the expected harmonics arise due to the frequency dependence of the permittivity. The resonance around 16 GHz arises from the lead system and is not related to the CPW. The inset shows a close-up of the fundamental resonance.

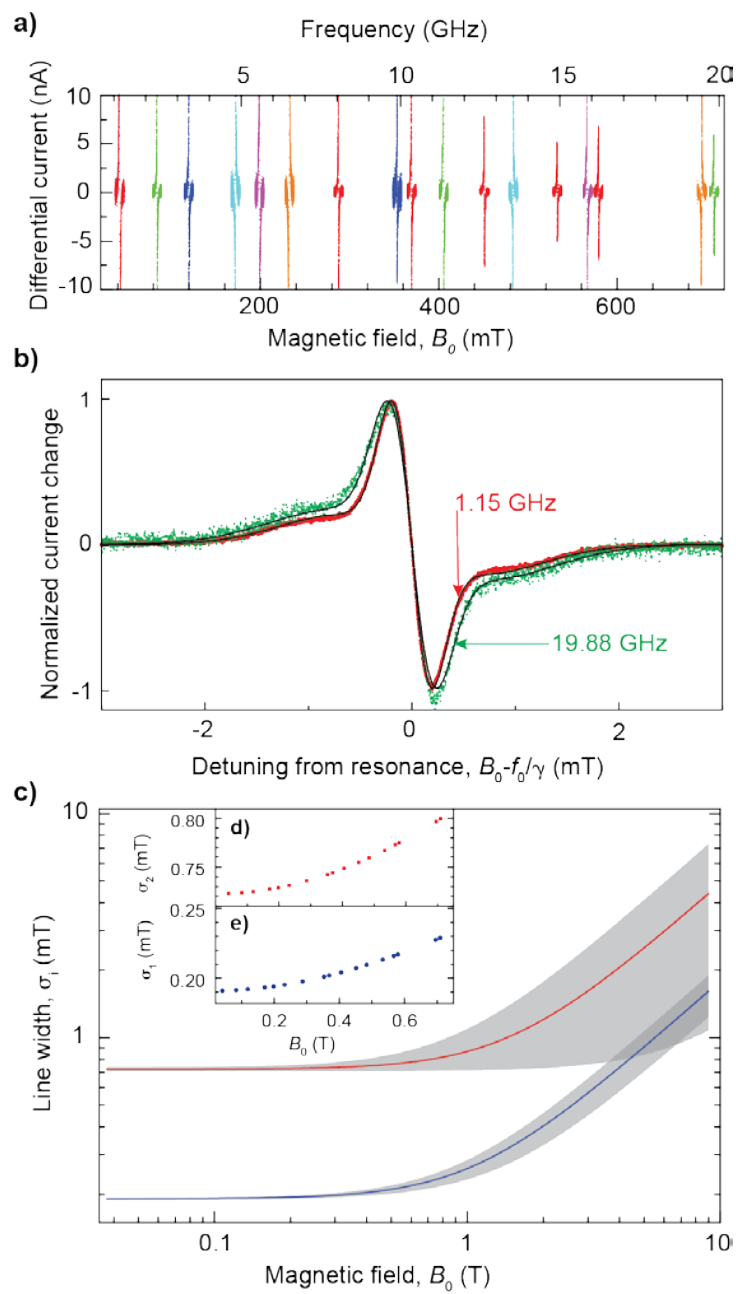

**Figure 3:** (color online) **Magnetic-field dependence of room-temperature OLED EDMR spectra. (a)** EDMR spectra obtained with six different CPW resonators (color coded) by utilizing the fundamental and some (but not all available) harmonic resonances. **(b)** Direct comparison of two normalized spectra at the lowest (1.15 GHz, red) and highest frequency (19.88 GHz, green), showing pronounced spectral broadening at high frequencies (corresponding to high $B_0$ fields). Black lines show fits to the measured spectra using

the derivative of the sum of two Gaussians, corresponding to electron and hole resonances. **(c)** Double-log plot of the widths of the two Gaussians as a function of $B_0$. A global fit routine of the individual resonance spectra is used to extrapolate the dependence up to 10 T (blue line for the narrow resonance, red line for the broad resonance) on a 95% confidence level (grey shaded area). The insets display points from the main plot in Fig. 3(c) corresponding to the magnetic fields at which the resonance spectra displayed in (a) were measured. These plots have linear axis scales and both abscissas have identical units. Note that these points do not represent raw data; they are values of the global fit of the EDMR spectra recorded at various microwave frequencies.